\documentclass[twocolumn]{svjour3}
\usepackage{graphicx}
\usepackage{caption}
\usepackage{subfig}
\renewcommand{\thefigure}{\arabic{figure}}

\renewcommand{\thetable}{\arabic{table}}

\begin{document}
\title{A two-species model of a two-dimensional sandpile surface: a case of asymptotic roughening}
\titlerunning{A two-species model of a two-dimensional sandpile surface} 
\author{Bandan Chakrabortty \and Anita Mehta}
\authorrunning{Bandan Chakrabortty \and Anita Mehta} 
\institute{
Bandan Chakrabortty \at 
S N Bose National Centre For Basic Sciences\\ 
Block-JD,Sector-III, Salt Lake, Kolkata 700098, India\\
\email{bandan@bose.res.in}           
\and
Anita Mehta \at
S N Bose National Centre For Basic Sciences\\ 
Block-JD,Sector-III, Salt Lake, Kolkata 700098, India\\
\email{anita@bose.res.in}
}
\date{Received: date / Accepted: date}
\maketitle
\begin{abstract}
We present and analyze a model of an evolving sandpile surface in ($2+1$) dimensions where the dynamics of  mobile grains ($\rho(\mathbf{x},t)$) and immobile clusters ($h(\mathbf{x},t)$) are coupled. Our coupling models the situation where the sandpile is flat on average, so that there is no bias due to gravity. We find anomalous scaling: the expected logarithmic smoothing at short length and time scales gives way to roughening in the asymptotic limit, where  novel and non-trivial exponents are found.
\end{abstract}

\section{Introduction}\label{introduction}

The study of sandpile surfaces unifies the field of granular physics \cite{am:book} with that of  surface growth \cite{krug:adv}, which latter has been the subject of prolonged theoretical analysis. One of the best known equations in surface growth is the Edwards-Wilkinson equation \cite{edwards:proc}, which has been used across a swathe of fields: from the experimental analysis of the epitaxial growth of thin films \cite{ac:jphysc} to the theoretical study of anomalous growth exponents in the presence of drift \cite{amjml:epl,pruessner:prl}. 

However, its use in the study of sandpile surfaces has been limited by its one-variable formulation, viz. the response of surface height $h(\mathbf{x},t)$ to the application of stochastic perturbations. Computer simulations of shaken granular packings in three dimensions \cite{am:prl} and cellular automaton models \cite{am:epl} suggested that a full description of the dynamics of granular surfaces needed the coupling of granular clusters \footnote{We use clusters to indicate collections of grains on surfaces which are relatively immobile} and flowing (mobile) grains. Since the growth and decay of clusters in a given region directly affects the surface height, it was thought appropriate \cite{am:book1} to couple the surface height $h(\mathbf{x},t)$, with a  variable $\rho(\mathbf{x},t)$ denoting the density of flowing grains across the surface. Equations were accordingly written down \cite{am:book1,am:pre} which embroidered the Edwards-Wilkinson equation with a variety of (usually) nonlinear transfer terms coupling $h(\mathbf{x},t)$ with  $\rho(\mathbf{x},t)$, reflecting different physical situations such as pouring/shaking a flat/sloping sandpile. The response to such perturbations could involve, for example, grains leaving their clusters and beginning to flow ($h(\mathbf{x},t)$ decreasing and $\rho(\mathbf{x},t)$ increasing at  ($\mathbf{x}, t$)): or a current of flowing grains getting lodged in a dip on the surface ($\rho(\mathbf{x},t)$ decreasing and $h(\mathbf{x},t)$ increasing at  ($\mathbf{x}, t$)). The purpose of the transfer term was to model such exchanges between the two species, and in so doing ensure the conservation of grains. A lively discussion ensued about the nature of these transfer terms, and many different ones were proposed to match evolving surfaces under different physical conditions \cite{bcre:prl,dg:1,dg:2,dg:3,dg:4}. Subsequently, the application of such noisy nonlinear two-species equations was extended to even more diverse situations, ranging from avalanching to ripple and dune formation \cite{aranson:pre,reb:prl,hans:pre}. 

All of the above approaches were unified by one particular feature: sandpile surfaces exhibited \textit{simple} scaling, independent of the length and time scales considered. However, experiments suggested that \textit{anomalous} scaling might arise \cite{sid}, with different exponents characterising short and long lengths/timescales. In particular, it was suggested that sloping sandpile surfaces might exhibit asymptotic smoothing under tilt \cite{sid1};  simulations suggested that this might be due to the large avalanches released, which `swept the surface clean' \cite{rough:pre}. New coupled equations (again drawing on the Edwards-Wilkinson equation \cite{edwards:proc}) were accordingly written down in one dimension, which explicitly investigated the effect of tilting the surface of a sloping sandpile: they showed that although roughening continued to be observed at short length and time scales, sandpiles manifested asymptotic smoothing \cite{pbis:pre}. The presence of bias is crucial to such smoothing, of course -- avalanches flow \textit{down}, rather than up a sloping sandpile. It was then natural to ask: what happens in the absence of such a bias, i.e. when a sandpile is \textit{flat} on average? Also, what happens in dimensions higher than one? The present paper asks, and then answers, these questions:  anomalous scaling is observed once again, but this time around, the crossover is from logarithmic smoothing for short length/time scales to asymptotic \textit{roughening}.

We now review some pertinent facts. In general, results of even single-species equations in more than one dimension, are hard to obtain analytically. An exception is the Edwards-Wilkinson equation \cite{edwards:proc}, which can be solved in any space dimensions ($d$): it is critical for $d=2$, with roughening exponents (to be defined below) $\alpha=2\beta=1-d/2$, where logarithmic scaling is observed \cite{krug:adv,tang:pra}. Even in the case of the KPZ equation \cite{kpz:prl} where an analytical solution exists for $d=1$ \cite{forster:pra,spohn:prl}, one has to resort to approximate numerical solutions and compare them with conjectures in $d=2$ \cite{forrest:jstat,wolf:epl,kim:prl}. 

We end this introductory section with a recap of scaling laws pertinent to surface growth \cite{healy:physrep,barabasi:book}. In general, power-law scaling is manifested by growing surfaces, so that  the interfacial width $W(L,t)$ of an initially flat interface continues to grow as a power of the time $t$ until it saturates: at saturation, its width  $W_{sat}(L,t)$ grows with a power of the system size $L$. Thus:
\begin{eqnarray}
W^{2}(L,t)\sim t^{2\beta},\label{scaling:relation1}\\
W^{2}_{sat}(L,t)\sim L^{2\alpha},\label{scaling:relation2}
\end{eqnarray}
where $\beta$ is the {\it growth exponent} and $\alpha$ is the {\it roughness exponent}. These two asymptotic power laws can be combined into a single scaling form,
\begin{equation}
W^{2}(L,t)\sim L^{2\alpha}f(t/L^{z}),
\end{equation}
with $z=\alpha/\beta$ the {\it dynamic exponent} and $f$ an appropriate scaling function. 

However, power-law scaling is sometimes replaced by \textit{logarithmic} scaling, especially for surfaces at their critical dimensionality \cite{krug:adv}. In such cases, we have:
\begin{eqnarray}
W^{2}(L,t)\sim \ln (t/a),\\ 
W^{2}_{sat}(L,t)\sim\ln (L/a),
\label{logscale}
\end{eqnarray}
where $a$ is a lattice constant. These two asymptotic expressions can be combined into a single scaling form \cite{tang:pra,hinrichsen:pre} to give:

\begin{equation}
{W^{2}(L,t)\sim\ln\left[\frac{L}{a}\ g\left(\frac{t}{L^{z}}\right)\right],}
\end{equation}
where $g$ is an appropriate scaling function. 

The plan of this paper is as follows: In Section \ref{equation} we present and describe our model equations; our results are presented and analysed  in Section \ref{numerical}. We discuss our results and make some concluding remarks in Section \ref{discussion}. 

\section{The model equations}\label{equation}

Our model equations in ($2+1$) dimensions contain a coupling between $h(\mathbf{x},t)$, the local height of immobile clusters and $\rho(\mathbf{x},t)$, the density of mobile particles; this is done through a transfer term $T$, as in previous approaches \cite{am:book}, and we follow closely the notation used earlier \cite{am:pre}:
\begin{eqnarray}
{\frac{\partial h(\mathbf{x},t)}{\partial t}=\nabla^{2}h(\mathbf{x},t)-T+\eta_{h}(\mathbf{x},t),\label{h_equation}}\\
{\frac{\partial \rho(\mathbf{x},t)}{\partial t}=\nabla^{2}\rho(\mathbf{x},t)+T+\eta_{\rho}(\mathbf{x},t),\label{rho_equation}}\\
T=-\rho(\mathbf{x},t)\nabla^{2}h(\mathbf{x},t)+(\nu-\mu\rho(\mathbf{x},t))|\nabla h(\mathbf{x},t)|\nonumber
\end{eqnarray}

The transfer term models exchanges between immobile clusters and flowing grains: the first part is proportional to surface diffusion, while the second part depends on the {\it magnitude} of the local slope at any point. This term is thus \textit{symmetric} with respect to the sign of the slope, and suggests that any deviation from a flat slope results in the conversion of grains from one to the other species. In more detail:

The first term $\rho(\mathbf{x},t) \nabla^{2}h(\mathbf{x},t)$, models the contribution to surface diffusion made by the `effective' surface of flowing grains when there is continuous avalanching ($\rho(\mathbf{x},t)\neq 0$ for all $\mathbf{x}$ and $t$) across the sandpile. The second term $\nu|\nabla h(\mathbf{x},t)|$  represents the spontaneous generation of flowing grains in response to (any direction of) tilt. The term $\mu\rho(\mathbf{x},t)|\nabla h(\mathbf{x},t)|$ represents the effect of a boundary layer below which the surface dynamics will not penetrate by limiting the release of flowing grains to this cutoff depth. The introduction of this term can be seen to provide a regulator, so that tilting will only ever be a moderate perturbation. After a short while when the condition $\langle\partial \rho(\mathbf{x},t) /\partial t\rangle =0 $ obtains, then the average value of $\rho(\mathbf{x},t)$ will be given by $\langle\rho(\mathbf{x},t)\rangle \simeq \nu/\mu$. This condition, apart from its physical reasonableness, is essential for numerical simulations, as it ensures that $h(\mathbf{x},t)$, $\rho(\mathbf{x},t)$ and their fluctuations remain finite and measurable. Thus $\mu$  plays the role of an experimental/numerical cut off \cite{jaeger:sci,am:rep} and the ratio of $\nu/\mu$  controls the inherent dynamics. 

The effects of shaking a sandpile (randomising the cluster distributions $h(\mathbf{x},t)$) and pouring grains on a sandpile (randomising the distribution of $\rho(\mathbf{x},t)$) are modelled by the noise terms $\eta_{h}(\mathbf{x},t)$ and $\eta_{\rho}(\mathbf{x},t)$ respectively. These are accordingly chosen to be two independent Gaussian white noises, characterised by their widths $\Delta_{h}$, $\Delta_{\rho}$ as follows:
\begin{eqnarray}
\langle\eta_{h}(\mathbf{x},t)\eta_{h}(\mathbf{x'},t')\rangle=\Delta^{2}_{h}\delta^{d}(\mathbf{x-x'})\delta(t-t'),\\
\langle\eta_{\rho}(\mathbf{x},t)\eta_{\rho}(\mathbf{x'},t')\rangle=\Delta^{2}_{\rho}\delta^{d}(\mathbf{x-x'})\delta(t-t').
\end{eqnarray}

\section{Numerical results and analysis}\label{numerical}

The simulations have been performed by discretizing Equation \ref{h_equation} and Equation \ref{rho_equation}, both in space and time. For this purpose we have implemented  multigrid algorithms \cite{briggs:multigrid} on a two-dimensional square lattice. Since it is critical behaviour that interests us, we have chosen relatively large spatial grids, ($a=\Delta x =\Delta y \sim  1.0$);  in order to avoid numerical instabilities, we have chosen relatively  small time stpng, $\Delta t \sim 0.001$. We have used a regulator function $F(T)$ to eliminate numerical divergences from the transfer term $T$ as in earlier work \cite{am:pre,pbis:pre}. This ensures that:
\begin{equation}
F(T)=\left\{\begin{array}{rcl}-1&\mbox{for}&T\leq -1\\T&\mbox{for}&-1\leq T\leq 1\\1&\mbox{for}&T\geq 1\end{array}\right.
\end{equation}
We start with the following intuition:
\begin{itemize}
\item Given the comments made in the introduction, we would expect the unperturbed behaviour of $h(\mathbf{x},t)$ to be logarithmically smooth, since the Edwards-Wilkinson equations has $d_c=2$.
\item We would likewise expect critical fluctuations  in $\rho(\mathbf{x},t)$ to be manifest only for $\nu\gg\mu$, because the mean value of $\rho(\mathbf{x},t)$ is constrained by the relation $\langle\rho(\mathbf{x},t)\rangle \simeq \nu/\mu$ (see above).
\end{itemize}

\begin{figure}[h]
\centering\includegraphics[width=4.0cm,height=4cm]{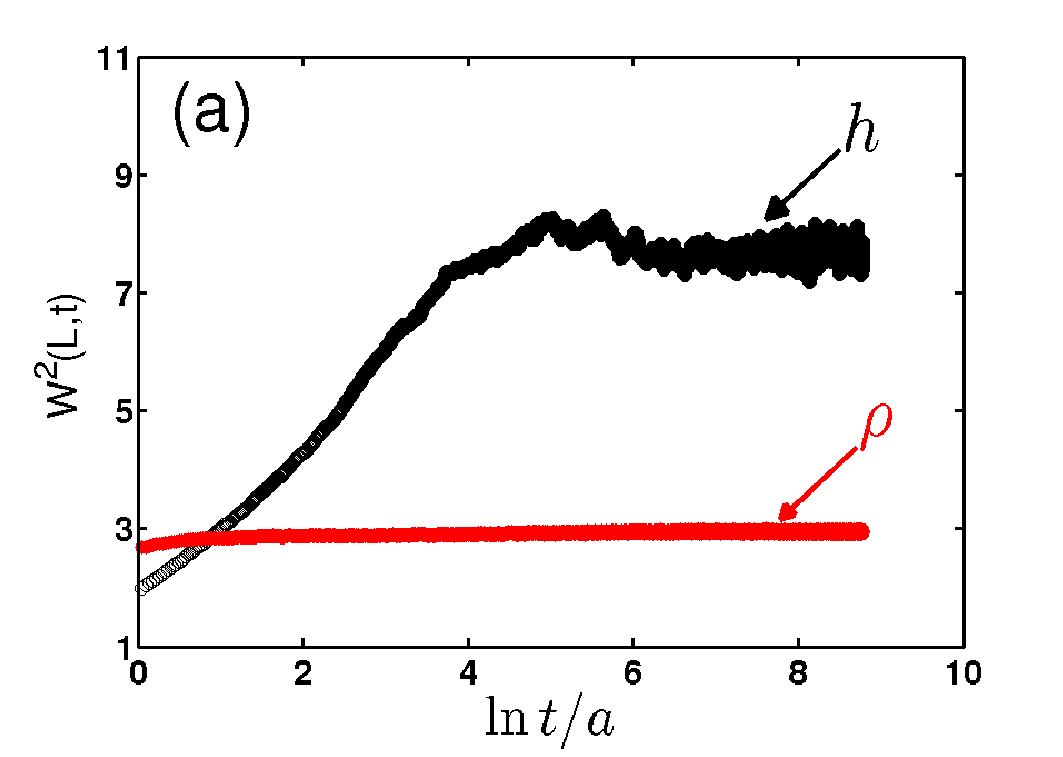}\hfill\includegraphics[width=4.0cm,height=4cm]{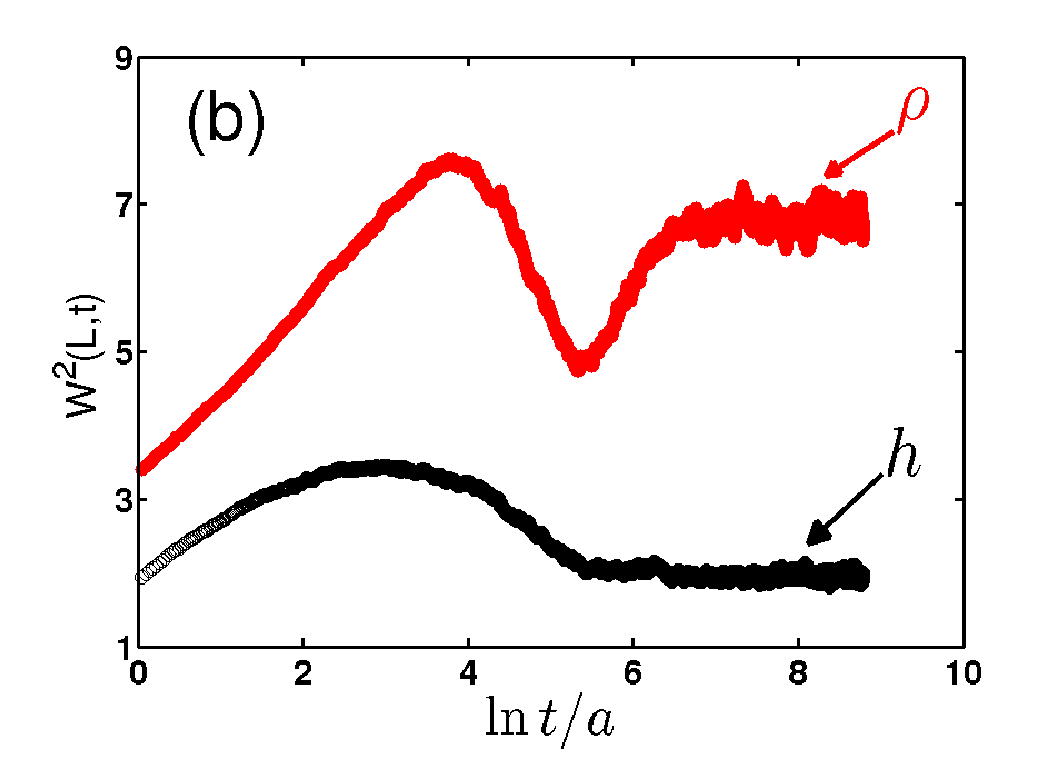}\\
\centering\includegraphics[width=4.1cm,height=4cm]{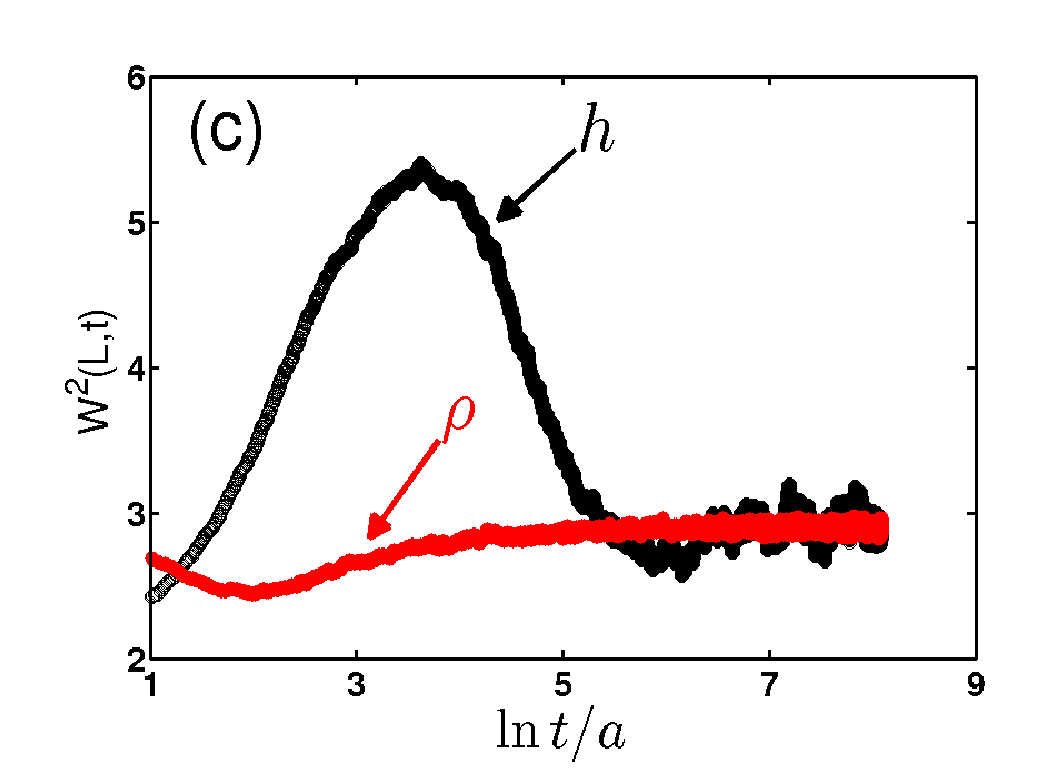}\hspace{5mm}\resizebox*{3.5cm}{3.6cm}{\includegraphics[trim=0cm -0.8cm 0cm 0cm, clip=true, angle=0]{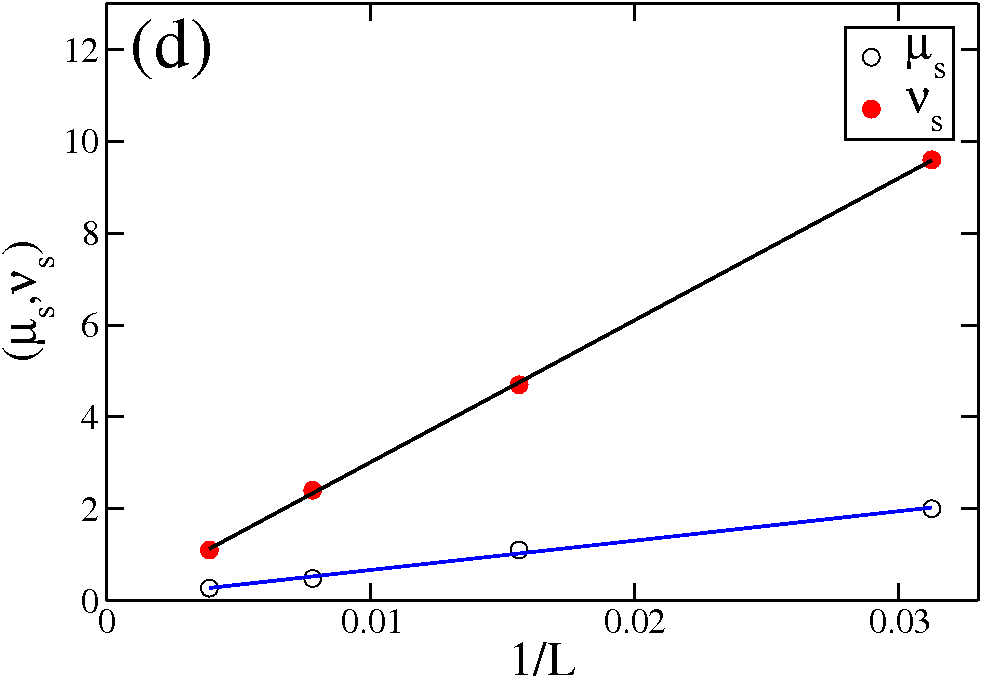}}
\caption{Square widths $W^{2}(L,t)$ are plotted  vs $\ln t/a$ with $L=64$ for both $h$ and $\rho$ species and for different values of $\mu,\nu$. Black indicates the $h$ species and the red  indicates the $\rho$ species in every graph. (a) A representative plot for $\mu \neq 0$ (chosen to be $\mu = 1.0$, in this case) and $\nu =0$. The qualitative behavior of the $h-\rho$ curves does not depend on the precise value of $\mu$ with in a large range. (b) The plot for $\mu =\nu = 0$. (c) When both $\mu$ and $\nu$ are non-zero, the fluctuations of $h$ and $\rho$ become comparable at a special point (see text). For $\mu = 1.0$, this occures at $\nu_{s}=4.7$; the corresponding graphs are presented. (d) The system size dependence of the coupling parameters at the special point ($\mu_s,\nu_s$) is explored by plotting ($\mu_s,\nu_s$) vs. $1/L$.}
\label{real}
\end{figure}

Before evaluating critical exponents, we examine the real-space behaviour of both species as a function of the relevant parameters.  According to the analytical predictions \cite{krug:adv,tang:pra} mentioned above,  we expect $W^{2}(L,t)$ to grow logarithmically with time for $d=2$, i.e $W^{2}(L,t)\sim\ln t/a$: accordingly, in Figure \ref{real}, we plot $W^{2}(L,t)$ vs $\ln t/a$ for different values of $\mu, \nu$. Although we have checked that our results do not depend on system size by checking for systems of size ($32\times 32, 64\times 64, 128\times 128, 256\times 256$), we present, for conciseness, only our results for $L=64$ in Fig. \ref{real}.  To begin with, in Figure \ref{real}a, we take $\nu=0$; our findings confirm the intuition that for any (non-zero) value of $\mu$, the $h(\mathbf{x},t)$ species is critical, while $\rho(\mathbf{x},t)$ has fully non-critical fluctuations. As soon as we set $\mu=0$, keeping $\nu=0$, we find -- surprisingly -- that the critical fluctuations of $\rho(\mathbf{x},t)$ begin to dominate those of $h(\mathbf{x},t)$ (Figure \ref{real}b).  In order to identify the point at which the critical fluctuations of both species become comparable, we fix $\mu=1.0$ and tune $\nu$: the results for this special point are shown in Figure \ref{real}c\footnote{The initial bump in these plots for $h(\mathbf{x},t)$ and $\rho(\mathbf{x},t)$ reflects the large initial value of $\rho(\mathbf{x},t)$ used in order to ensure the positivity of $\rho(\mathbf{x},t)$  throughout; this however is not expected to affect the long-time scaling behaviour, as our plots confirm}. We examine the system size dependence of this special point by
analysing systems of size $32\times 32, 64\times 64, 128\times 128, 256\times 256$; the critical behaviour of both species is unchanged if we rescale the values of $\mu$ and $\nu$ with system size. This rescaling is made explicit in Figure \ref{real}d, which suggests these special values $\mu_s$ and $\nu_s$ depend systematically on system size as $\sim 1/L$. 

In order to evaluate the critical exponents, we evaluate the structure factors $S(t_{sat},\kappa)$ and $S(L,\omega)$, which are related to the interfacial width $W^{2}(L,t)$  as follows \cite{barabasi:book}:\\
\begin{equation}
W^{2}(L,t)=\frac{1}{L^{d}}\sum\limits_{\kappa}S(\kappa ,t)=\frac{1}{t^{d}}\sum\limits_{\omega}S(L ,\omega).
\end{equation}
Because of the above relation,
the scaling relations Equation \ref{scaling:relation1} and Equation \ref{scaling:relation2} can be recast as,
\begin{eqnarray}
S(t_{sat},\kappa)\sim \kappa^{-2-2\alpha}\ (\kappa \rightarrow 0),\\
S(L,\omega)\sim \omega^{-2-2\beta}\ (\omega \rightarrow 0),
\end{eqnarray}
here, $t_{sat}$ is the saturation time of the interfacial width $W(L,t)$ and  the wave vector $\kappa =|\mathbf{\kappa}|=\sqrt{\kappa_{x}^{2}+\kappa_{y}^{2}}$.

We mention first of all that the critical behavior of the Edwards-Wilkinson equation in 2d is logarithmic ($\alpha =\beta =0$). In the work of this paper, crossovers to different nontrivial exponents are manifested as a function of different parameters. In particular, we emphasizes that {\it all} the nontrivial scaling behaviors found here are {\it anomalous}, i.e. shows different scaling for long length and time scales. 

We proceed by first evaluating the cases considered in Figure \ref{real}, corresponding to the physical situation where there is noise in both species (i.e. $\Delta^{2}_{h}=\Delta^{2}_{\rho}>0$). The plots of structure factor for each set of parameter values considered are presented successively, while the exponent values are compiled and tabulated in Table \ref{ta} and Table \ref{tb} (see below).

The first case is presented in Figure  \ref{mu4nu0}, for the parameter values $\nu=0$ and $\mu\neq 0$. Based on Figure  \ref{real}a, we would not expect to see nontrivial critical fluctuations for $\rho(\mathbf{x},t)$, an expectation borne out by Table \ref{ta} and Table \ref{tb}. Physically, this is because the absence of tilt ($\nu =0$) halts the spontaneous generation of mobile grains in the steady state, when $<\rho(\mathbf{x},t)>\simeq 0$; the only source of flowing grains is $\eta_{\rho}(\mathbf{x},t)$.

The  case of $\nu= \mu =0$ is examined in Figure \ref{mu0nu0}, where Figure \ref{real}b would lead us to expect that the fluctuations of $\rho(\mathbf{x},t)$ would dominate those of $h(\mathbf{x},t)$; a perusal of  Table \ref{ta} and Table \ref{tb} confirms that this is the case. In this case, the dynamical equation for $h(\mathbf{x},t)$ can be viewed as a linear Edwards-Wilkinson equation having an effective diffusion constant of ($1+\rho(\mathbf{x},t)$).
The special point ($\mu_s, \nu_s$) characterises the crossover point where the fluctuations of $\rho(\mathbf{x},t)$ just begin to become `dangerous', i.e. critical, and of the same order as those of $h(\mathbf{x},t)$. The plots of
Figure \ref{cri} which lead to the exponents tabulated in Table \ref{ta} and Table \ref{tb} confirm this intuition: both species show nearly comparable asymptotic roughening.

\begin{figure}[h]
\centering\subfloat[Coupling parameters used were:  $\nu = 0.0,\mu
 = 4.0$.]{\includegraphics[trim=0cm 0cm 0cm 0.1cm, clip=true,width=4.0cm,height=3.5cm]{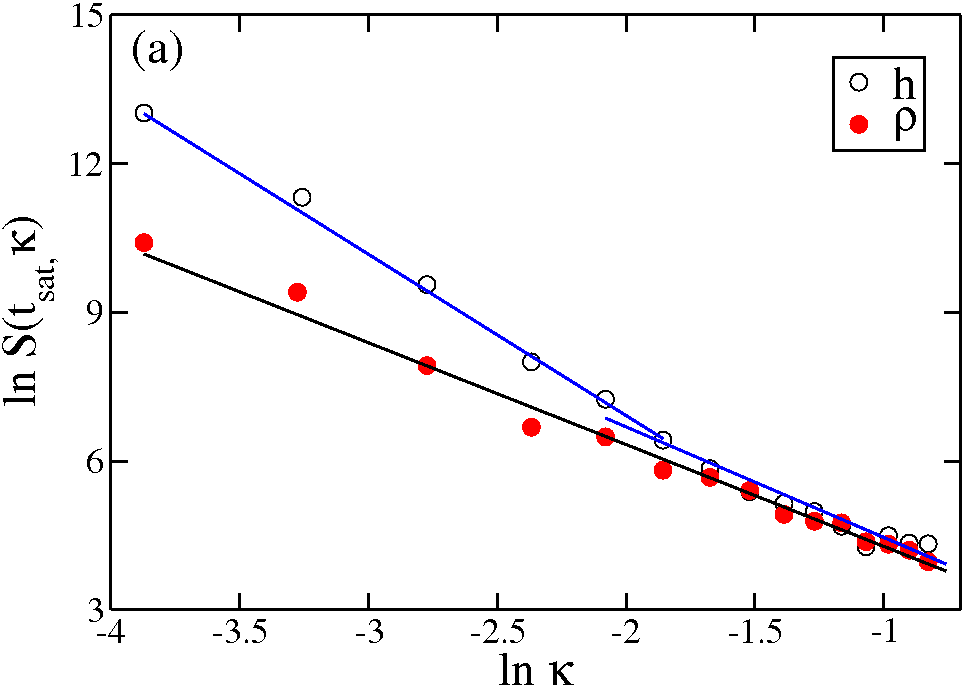}\hspace{0.2cm}\includegraphics[trim=0cm 0cm 0cm 0.1cm, clip=true,width=4.0cm,height=3.5cm,width=4.0cm,height=3.5cm]{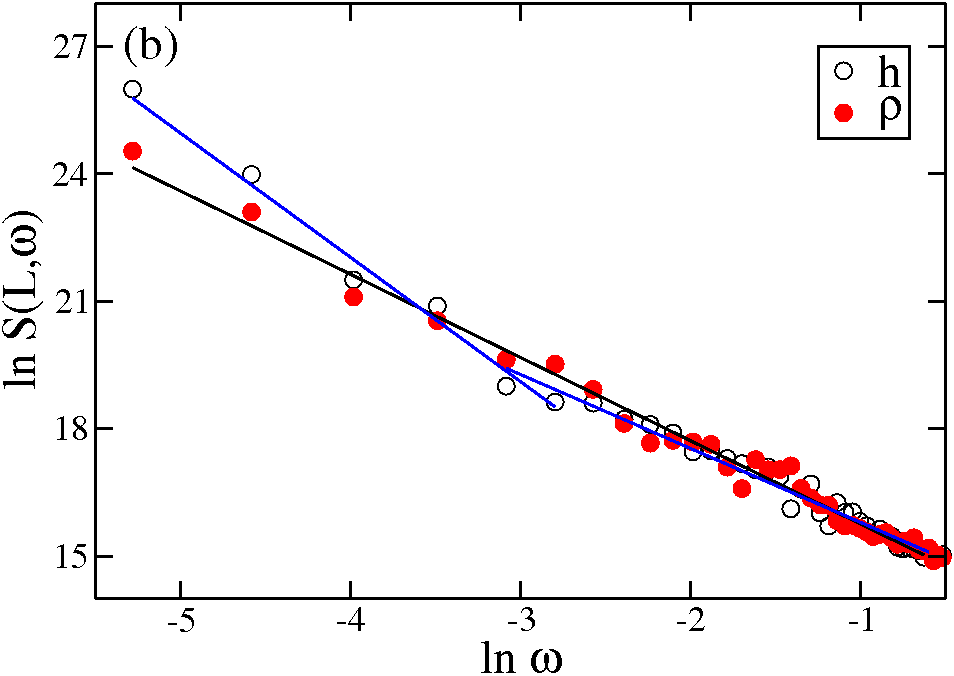}\label{mu4nu0}}\\
\vspace{0.3cm}
\centering\subfloat[Coupling parameters used were: $\nu = 0.0,\mu = 0.0$.]{\includegraphics[width=4.0cm,height=3.5cm]{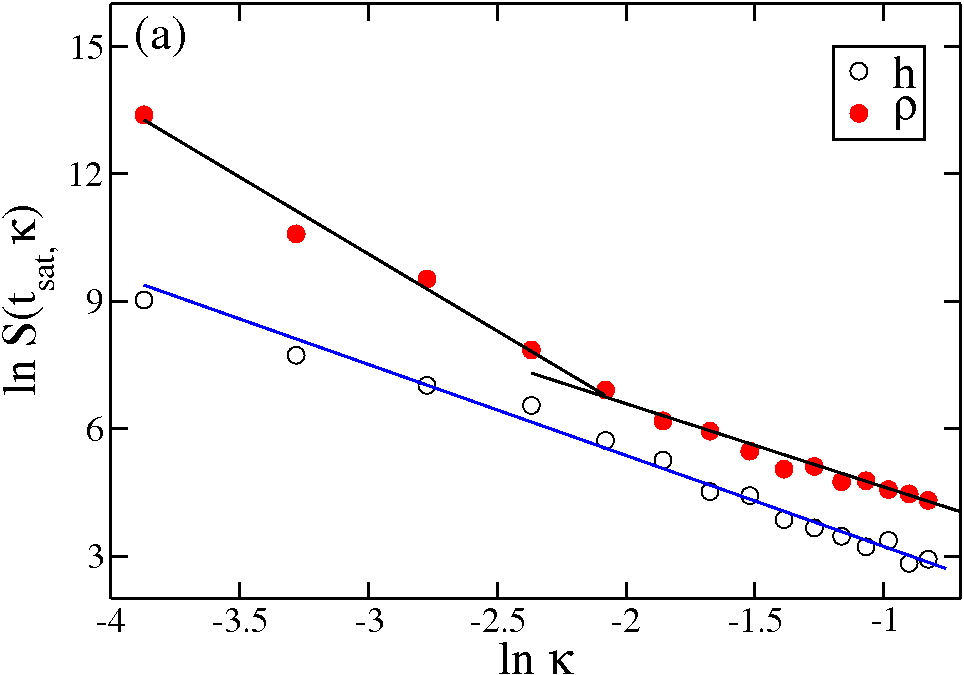}\hspace{0.2cm}\includegraphics[width=4.0cm,height=3.5cm]{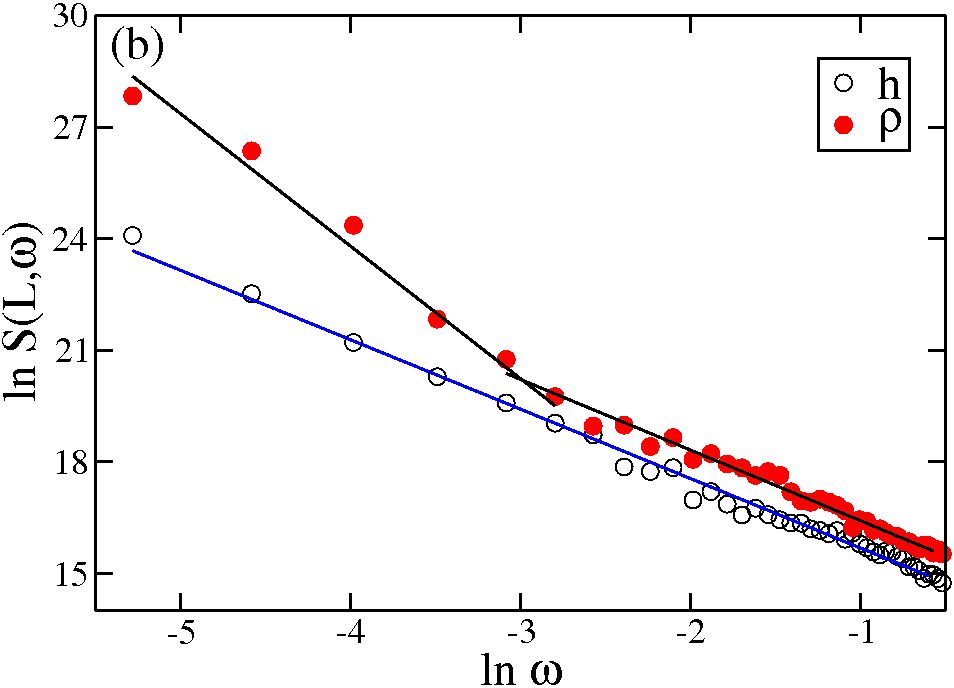}\label{mu0nu0}}\\
\vspace{0.3cm}
\centering\subfloat[Coupling parameters used were: $\nu_{s} = 4.7,\mu_{s} = 1.0$. ]{\includegraphics[width=4.0cm,height=3.5cm]{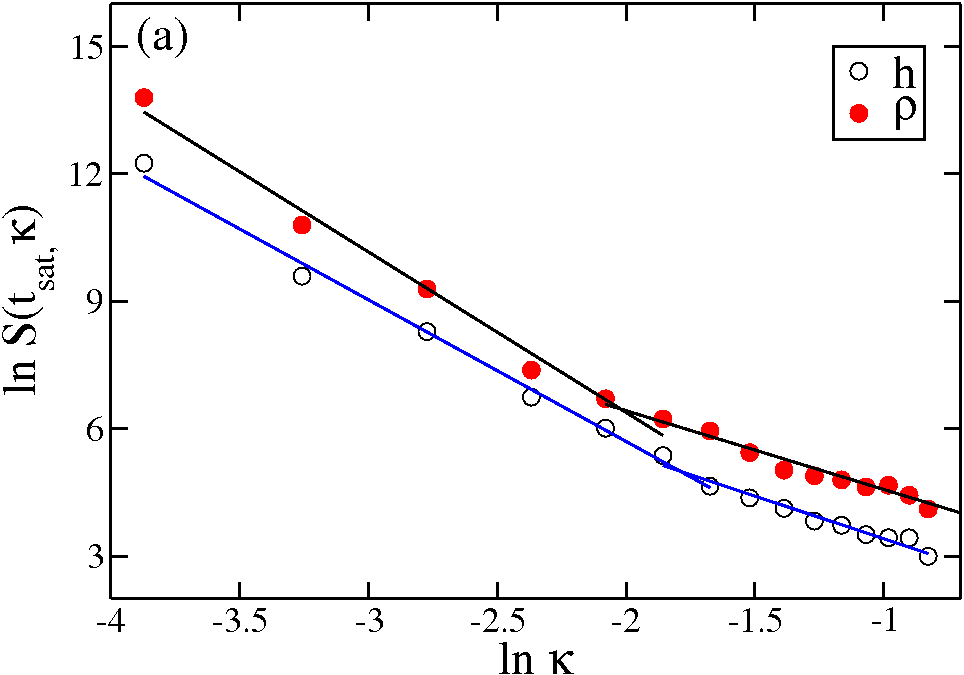}\hspace{0.2cm}\includegraphics[width=4.0cm,height=3.5cm]{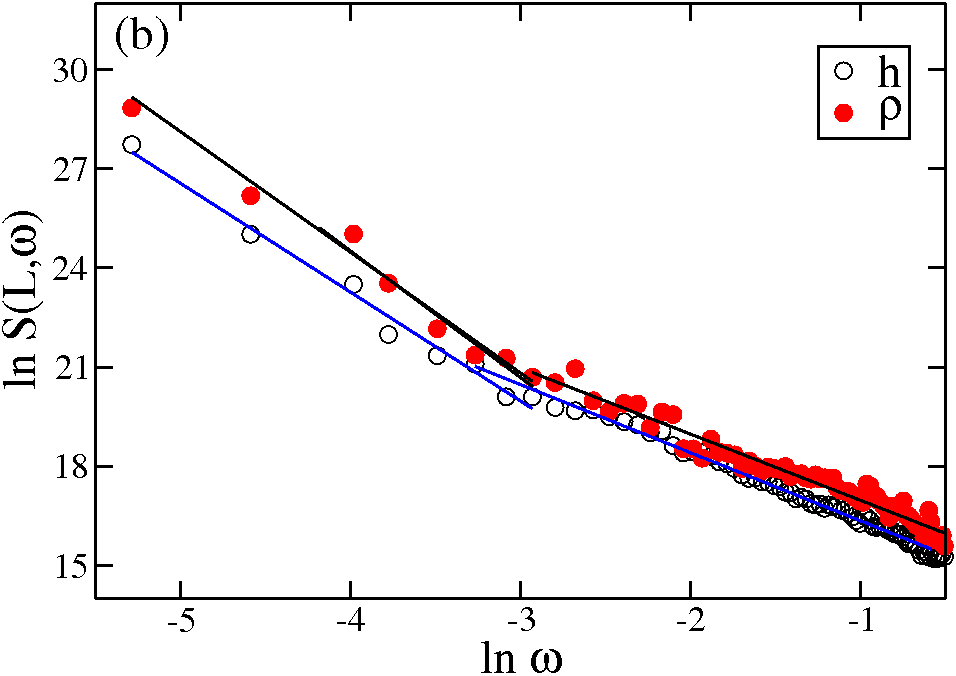}\label{cri}}\\
\vspace{0.3cm}
\centering\subfloat[Coupling parameters used were: $\nu = 15.0,\mu = 1.0$.]{\includegraphics[width=4.0cm,height=3.5cm]{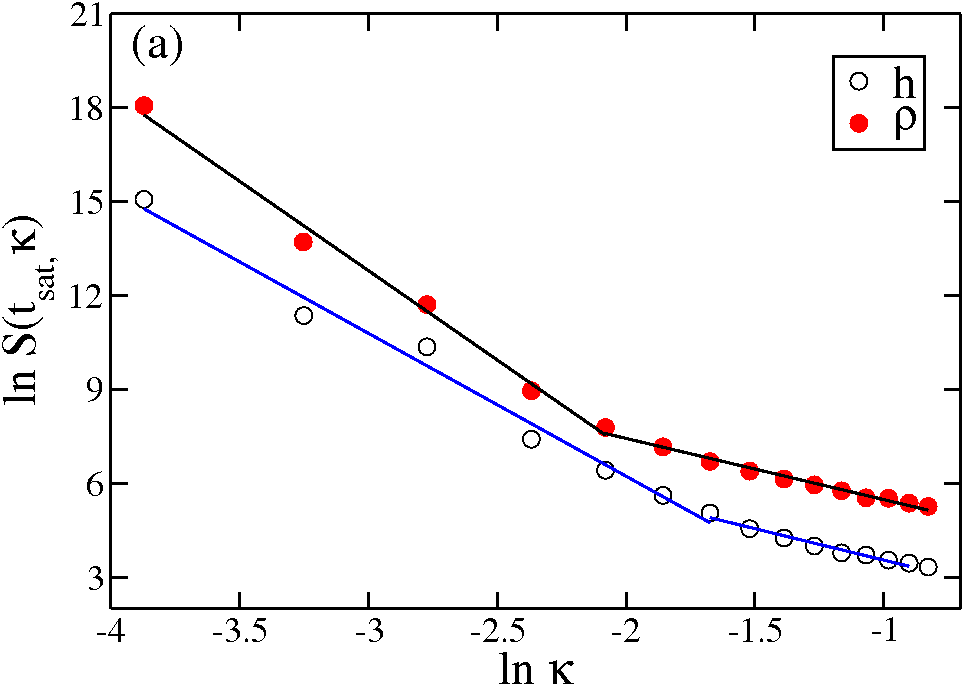}\hspace{0.2cm}\includegraphics[width=4.0cm,height=3.5cm]{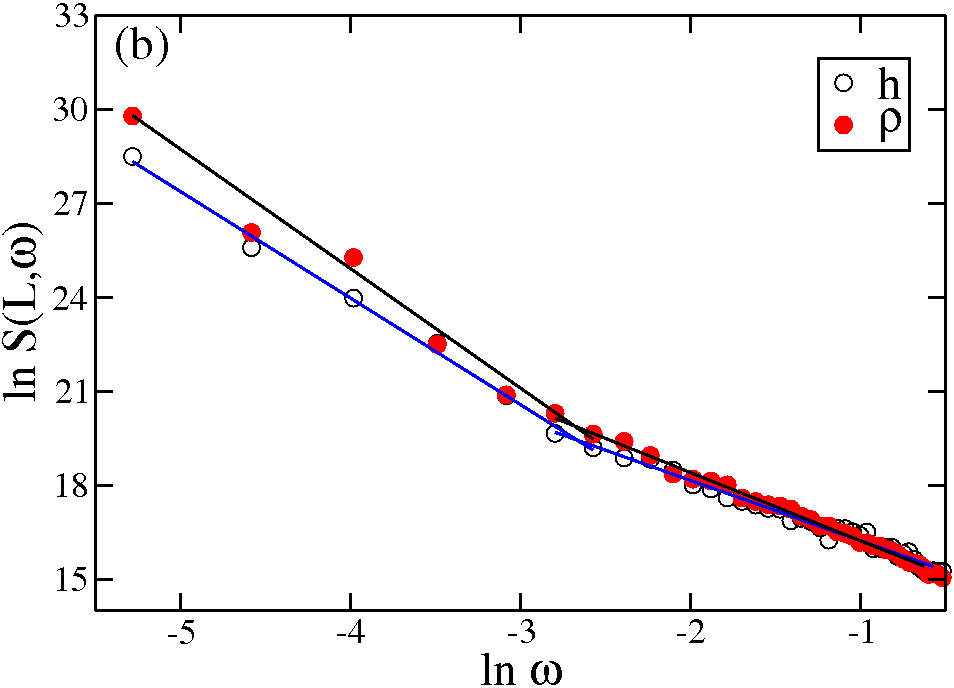}\label{mu1nu15}}
\caption{ Plots of structure factors for the different cases (in the presence of noise in both the species:  $\Delta^{2}_{h}=\Delta^{2}_{\rho}=1.0$\label{sf:noise}): (a) $\ln S(t_{sat},\kappa)$ vs $\ln \kappa$  plot and (b) $\ln S(L,\omega)$ vs $\ln \omega$ plot. }
\end{figure}

\newpage
Above the special point ($\mu_s$ fixed, $\nu > \nu_s$), we would expect that the fluctuations of $\rho(\mathbf{x},t)$ would dominate those of $h(\mathbf{x},t)$ (as in the case of Figure \ref{mu0nu0}); our results, presented in Figure \ref{mu1nu15}, and tabulated in  Table \ref{ta} and Table \ref{tb}, confirm this. This can be understood intuitively by realising that for the $\rho(\mathbf{x},t)$ equation (and when $\rho(\mathbf{x},t)$ is finite), the term $\rho(\mathbf{x},t) \nabla^{2}h$ \textit{cannot} be renormalised away as an addition to an effective diffusion constant, as it can in the case of the $h(\mathbf{x},t)$ equation.

\begin{figure}[h]
\centering\includegraphics[width=4.0cm,height=3.6cm]{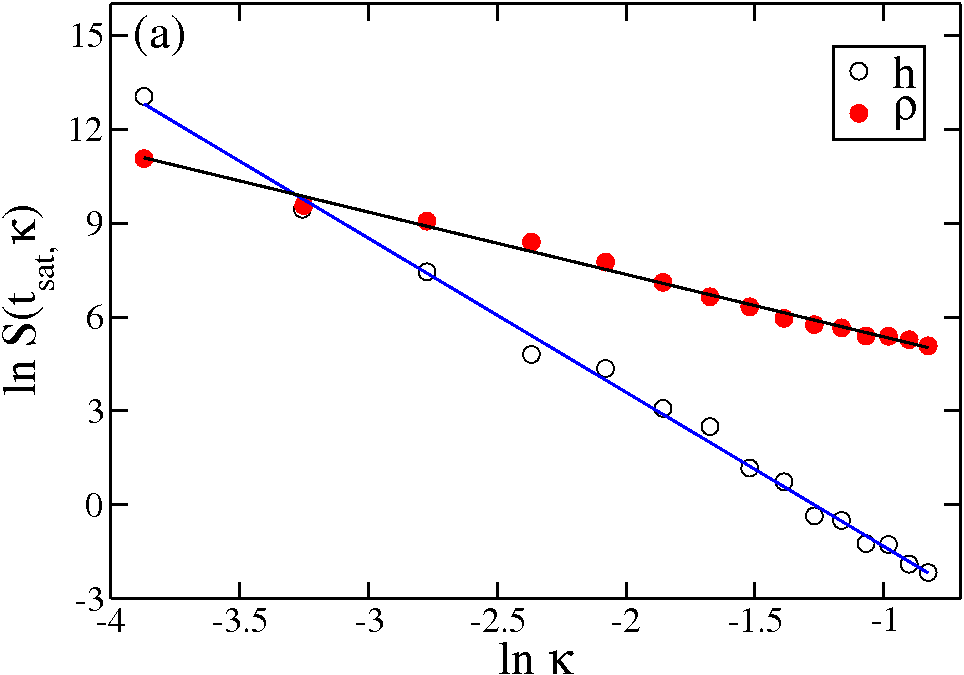}\hfil\includegraphics[width=4.0cm,height=3.6cm]{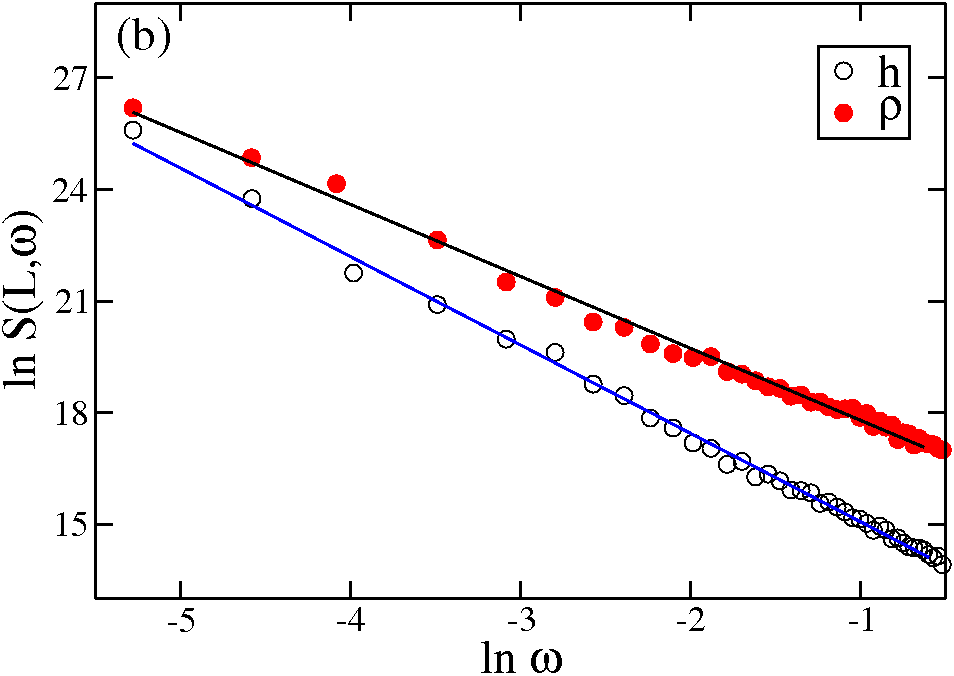}
\caption{ Plots of structure factors in the absence of noise in $h$ species ($\Delta^{2}_{h}=0$ after a transient and $\Delta^{2}_{\rho}>0$): (a) $\ln S(t_{sat},\kappa)$ vs $\ln \kappa$ plot and (b) $\ln S(L,\omega)$ vs $\ln \omega$ plot. The coupling parameters used were: $\nu = 15.0, \mu =1.0$ but the results hold for arbitrary values.\label{sp}}
\end{figure}

Finally, we study the situation when there is only sustained noise in $\rho(\mathbf{x},t)$: i.e. $\Delta^{2}_{h}=0$ after a transient and $\Delta^{2}_{\rho}>0$. From a study of the model equations in one dimension under these conditions \cite{am:pre}, we would expect that the $h(\mathbf{x},t)$ landscape would freeze up after a transient (with possibly non-trivial roughening exponents), while the $\rho(\mathbf{x},t)$ landscape would diffuse over this frozen landscape with Edwards-Wilkinson exponents. We find indeed (Figure \ref{sp}, Table \ref{ta} and Table \ref{tb}) that after a transient, the $h(\mathbf{x},t)$ landscape is frozen, with non-trivial roughening exponents entirely inherited from the transient noise in $h(\mathbf{x},t)$. At this point, the $\rho(\mathbf{x},t)$ landscape gets decoupled from the $h(\mathbf{x},t)$ landscape, and shows logarithmic roughening.

\begin{table}[h]
\vspace{0.5cm}
\setlength{\tabcolsep}{1pt}
\centering\subfloat[Values of $\alpha$.\label{ta}]{
\begin{tabular}{|c|c|c|c|}
\hline
Mode of coupling&Species&\multicolumn{2}{c|}{$\alpha$}\\
\cline{3-4}
&& for large $\kappa$&for small $\kappa$\\
\hline
Edwards-Wilkinson&$h(\mathbf{x},t)$&$0.0$&$0.0$\\
\hline
$\nu=0,\mu\neq 0$&$h(\mathbf{x},t)$&$0.00\pm0.05$&$0.62\pm 0.06$\\
$\Delta^{2}_{h}=\Delta^{2}_{\rho}>0$&$\rho(\mathbf{x},t)$&$-0.02\pm0.04$&$-0.02\pm 0.04$\\
\hline
$\nu=0,\mu= 0$&$h(\mathbf{x},t)$&$0.07\pm0.08$&$0.07\pm 0.08$\\
$\Delta^{2}_{h}=\Delta^{2}_{\rho}>0$&$\rho(\mathbf{x},t)$&$0.00\pm0.05$&$0.81\pm 0.07$\\
\hline
$\nu_s,\mu_s (special),\nu_s>\mu_s $&$h(\mathbf{x},t)$&$0.00\pm0.03$&$0.66\pm 0.06$\\
$\Delta^{2}_{h}=\Delta^{2}_{\rho}>0$&$\rho(\mathbf{x},t)$&$0.00\pm 0.06$&$0.85\pm 0.05$\\
\hline
$\nu,\mu (> \nu_s,\mu_s), \nu >\mu $&$h(\mathbf{x},t)$&$0.00\pm0.05$&$1.28\pm 0.04$\\
$\Delta^{2}_{h}=\Delta^{2}_{\rho}>0$&$\rho(\mathbf{x},t)$&$0.00\pm0.04$&$1.85\pm 0.06$\\
\hline
arbitrary $\mu ,\nu$&$h(\mathbf{x},t)$&$1.45\pm0.08$&$1.45\pm 0.08$\\
$\Delta^{2}_{h}=0,\Delta^{2}_{\rho}>0$&$\rho(\mathbf{x},t)$&$0.00\pm0.04$&$0.00\pm 0.04$\\
\hline
\end{tabular}
}\\
\centering\subfloat[Values of $\beta$.\label{tb}]{
\begin{tabular}{|c|c|c|c|}
\hline
Mode of coupling&Species&\multicolumn{2}{c|}{$\beta$}\\
\cline{3-4}
&& for large $\omega$&for small $\omega$\\
\hline
Edwards-Wilkinson&$h(\mathbf{x},t)$&$0.0$&$0.0$\\
\hline
$\nu=0,\mu\neq 0$&$h(\mathbf{x},t)$&$0.00\pm0.03$&$0.50\pm 0.04$\\
$\Delta^{2}_{h}=\Delta^{2}_{\rho}>0$&$\rho(\mathbf{x},t)$&$-0.03\pm0.04$&$-0.03\pm 0.04$\\
\hline
$\nu=0,\mu= 0$&$h(\mathbf{x},t)$&$0.01\pm0.05$&$0.01\pm 0.05$\\
$\Delta^{2}_{h}=\Delta^{2}_{\rho}>0$&$\rho(\mathbf{x},t)$&$0.00\pm0.04$&$0.76\pm 0.06$\\
\hline
$\nu_s,\mu_s (special), \nu_s>\mu_s$&$h(\mathbf{x},t)$&$0.00\pm0.04$&$0.64\pm 0.02$\\
$\Delta^{2}_{h}=\Delta^{2}_{\rho}>0$&$\rho(\mathbf{x},t)$&$0.00\pm0.06$&$0.76\pm 0.04$\\
\hline
$\nu,\mu (> \nu_s,\mu_s),  \nu>\mu$&$h(\mathbf{x},t)$&$0.00\pm0.05$&$0.72\pm 0.05$\\
$\Delta^{2}_{h}=\Delta^{2}_{\rho}>0$&$\rho(\mathbf{x},t)$&$0.00\pm0.08$&$0.85\pm 0.07$\\
\hline
arbitrary $\mu ,\nu$&$h(\mathbf{x},t)$&$0.16\pm0.03$&$0.16\pm 0.03$\\
$\Delta^{2}_{h}=0,\Delta^{2}_{\rho}>0$&$\rho(\mathbf{x},t)$&$0.00\pm0.07$&$0.00\pm 0.07$\\
\hline
\end{tabular}
}
\caption{Numerically estimated values of $\alpha$ and $\beta$, for the various physical situations described in this paper.}
\end{table}

Before leaving this section, we summarise our findings: except when the tilt term vanishes, the critical fluctuations of $\rho(\mathbf{x},t)$ dominate those of $h(\mathbf{x},t)$ as long as there is noise in both species $\Delta^{2}_{h}>0,\Delta^{2}_{\rho}>0$. When, however, $\Delta^{2}_{h}=0$ after a transient and $\Delta^{2}_{\rho}>0$,  the $\rho(\mathbf{x},t)$ species is logarithmically smooth, while the frozen $h(\mathbf{x},t)$ landscape is characterised by a novel non-trivial roughening exponent, entirely inherited from the transient phase.

\section{Discussion}\label{discussion}
The main focus of our paper was the investigation of a two-species model of sandpile dynamics, representing the height profile and the dynamics of granular flow along a flat surface in the presence of perturbations such as shaking and/or pouring. Importantly, no bias exists in this system, so that avalanches of grains can flow in both directions: this is to be contrasted with the downward flow of grains due to the biasing field of gravity, on a sloping sandpile. Given that the latter scenario had led to asymptotic smoothing \cite{rough:pre,pbis:pre}, we wanted to investigate the effect of making our exchange terms fully symmetric, to model a flat sandpile surface.

The particular model that we have explored has an exchange term that includes the effect of tilt, as well as that of a finite boundary layer, below which
the bulk of the sandpile is protected from the dynamics at the surface. A major difference from the investigation of similar equations in one dimension \cite{am:pre} is that \textit{anomalous} scaling is obtained in general for the most \textit{relevant} species: wherever appropriate, short-wavelength logarithmic smoothing behaviour at small wavelengths crosses over to asymptotic \textit{roughening}. This is largely due to the action of the tilt term, which in this unbiased case, causes flowing grains to be generated independent of the sign of the local slope: dips and bumps alike accumulate flowing grains,  a change from the asymmetric (biased) situation considered in \cite{pbis:pre}.

We now discuss our results in detail. When $\nu = 0$, flowing grains are not generated from clusters; the only source for these is from the `pouring' source, $\eta_{\rho}(\mathbf{x},t)$. The form of the transfer term then implies that there is a net transfer from the flowing grains to the clusters, resulting in an enhanced roughness as the couplings become relevant; the $h(\mathbf{x},t)$ species is then characterised by a crossover from logarithmic smoothing to roughening with the novel exponents presented in the second line of each of Tables \ref{ta} and \ref{tb} , while the $\rho(\mathbf{x},t)$ species remains logarithmically smooth as expected. Such a simplification also occurs (sixth line of Tables \ref{ta} and \ref{tb}) when after a transient period when both noises are present, only  $\Delta^{2}_{\rho}$ remains finite. After a transient period, the $h(\mathbf{x},t)$ profile is frozen into a state of extreme roughness, fed by the flowing grains (whose dynamics show the expected EW logarithmic smoothing): the large roughening exponent obtained is novel, especially because it is independent of parameter values. Finally, for $\nu>\mu$ such that $(\nu/\mu)\geq (\nu_s/\mu_s)$, the critical fluctuations of $\rho(\mathbf{x},t)$ dominate those of $h(\mathbf{x},t)$, which can be viewed as a state of continuous avalanching, when the effective surface is defined by the film of flowing grains \cite{pbis:pre}. Anomalous scaling is again everywhere observed, with a crossover from logarithmic smoothing to asymptotic roughening: this can be given an interpretation in the context of `thick avalanches' \cite{dg:4}, so that the novel roughening exponents we have reported for $\rho(\mathbf{x},t)$  can be analysed in greater depth with respect to earlier studies \cite{dg:3}. They can also be measured using sophisticated visualisation techniques on traditional rotating cylinder experiments, which have been the subject of considerable recent activity \cite{hill:gm}, and which can fruitfully be employed in studies of the kind of surface roughening reported here.

\begin{acknowledgements}
BC would like to thank the University Grants Commission (UGC), India, for financial support. We thank Dr J M Luck for very helpful discussions and a critical reading of the manuscript.
\end{acknowledgements}

\end{document}